\documentclass[Physsubmission, Phys]{SciPost}

\hypersetup{
    colorlinks,
    linkcolor={red!50!black},
    citecolor={blue!50!black},
    urlcolor={blue!80!black}
}

\usepackage[bitstream-charter]{mathdesign}
\DeclareSymbolFont{usualmathcal}{OMS}{cmsy}{m}{n}
\DeclareSymbolFontAlphabet{\mathcal}{usualmathcal}

\usepackage{bm}
\usepackage{braket}

\begin{document}

\begin{center}{\Large \textbf{
Confinement, mass gap 
and gauge symmetry in the Yang-Mills theory
-- restoration of residual local gauge symmetry --
}}\end{center}

\begin{center}
Kei-Ichi Kondo\textsuperscript{1$\star$} and
Naoki Fukushima\textsuperscript{2} 
\end{center}

\begin{center}
{\bf 1} Department of Physics, Graduate School of Science, Chiba University, Chiba 263-8522, Japan
\\
{\bf 2} Department of Physics, Graduate School of Science and Engineering, Chiba University, Chiba 263-8522, Japan
\\
* 
kondok@faculty.chiba-u.jp
\end{center}

\begin{center}
\today
\end{center}


\definecolor{palegray}{gray}{0.95}
\begin{center}
\colorbox{palegray}{
  \begin{minipage}{0.95\textwidth}
    \begin{center}
    {\it  XXXIII International (ONLINE) Workshop on High Energy Physics \\“Hard Problems of Hadron Physics:  Non-Perturbative QCD \& Related Quests”}\\
    {\it November 8-12, 2021} \\
    \doi{10.21468/SciPostPhysProc.?}\\
    \end{center}
  \end{minipage}
}
\end{center}

\section*{Abstract}
{\bf
In this talk we want to discuss the color confinement criterion which guarantees confinement of all colored particles including dynamical quarks and gluons. 
The most well-known criterion is the Kugo-Ojima color confinement criterion derived in the Lorenz gauge. 
However, it was pointed out that the Kugo-Ojima criterion breaks down for the Maximal Abelian gauge in which quark confinement has been verified according to the dual superconductivity caused by magnetic monopole condensations.  
We give the color confinement criterion based on the restoration of the residual local gauge symmetry which can be applied to the Abelian and non-Abelian gauge theories as well irrespective of the compact or non-compact formulation, 
and enables us to understand confinement in all the cases. 
Indeed, the restoration of the residual local gauge symmetry which was shown by Hata in the Lorenz gauge to be equivalent to the Kugo-Ojima criterion  indeed occurs  in the Maximal Abelian gauge for the SU(N) Yang-Mills theory in two-, three- and four-dimensional Euclidean spacetime once the singular topological configurations of gauge fields are taken into account. 
This result indicates that the color confinement phase is a disordered phase caused by non-trivial topological configurations irrespective of the gauge choice. 
}


\section{Introduction}
\label{sec:intro}

Quark confinement is well understood based on the dual superconductor picture \cite{dualsuper} where condensation of magnetic monopoles and antimonopoles occurs. 
For a review, see e.g. \cite{CP97} and \cite{KKSS15}. 
Even if the dual superconductor picture is true,  however, it is not an easy task to apply this picture to various composite particles composed of quarks and/or gluons. 
In fact,  gluon confinement is still less understood, although there are interesting developments quite recently, see \cite{HK21} and reference therein. 

In view of these, we recall the color confinement due to Kugo and Ojima (1979) \cite{KO79}. 
If the Kugo and Ojima (KO) criterion is satisfied, all colored objects cannot be observed.  
Then quark confinement and gluon confinement immediately follow as special cases of color confinement.

However, the KO criterion was derived only in the Lorenz gauge $\partial^\mu \mathscr{A}_\mu=0$, even if the issue on the existence of the nilpotent BRST symmetry is put aside for a while. 

The KO criterion is written in terms of a specific correlation function called the KO function which is clearly gauge-dependent and is not directly applied to the other gauge fixing conditions. 

From this point of view, the maximally Abelian (MA) gauge \cite{MAG} 
is the best gauge to be investigated because the dual superconductor picture for quark confinement was intensively investigated in the MA gauge. 

Nevertheless, 
Suzuki and Shimada (1983) \cite{SS83} 
 pointed out that the KO criterion cannot be applied to the MA gauge and the KO criterion is violated in the model for which quark confinement is shown to occur by Polyakov (1977) \cite{Polyakov77} 
due to magnetic monopole and antimonopole condensation.  
%
 Hata and Niigata (1993) \cite{HN93}
  claimed that the MA gauge is an exceptional case to which the KO color confinement criterion cannot be applied. 

We wonder how the color confinement criterion of the KO type is compatible with the dual superconductor picture for quark confinement.

We reconsider the color confinement criterion of the KO type in the Lorenz gauge and give an explicit form to be satisfied in the MA gauge within the same framework as the Lorenz gauge in the manifestly Lorentz covariant operator formalism with the unbroken BRST symmetry \cite{KF21}.

For this purpose, we make use of the method of Hata (1982) \cite{Hata82}
 claiming that 
the KO criterion is equivalent to the condition for the residual local gauge symmetry to be restored. 
The usual gauge fixing condition is sufficient to fix the gauge in the perturbative framework in the sense that it enables us to perform perturbative calculations.
However, it  does not eliminate the gauge degrees of freedom entirely but leaves certain gauge symmetry which is  called the \textbf{residual local gauge symmetry}.  The residual local gauge symmetry can in principle be spontaneously broken.  
This phenomenon does not contradict the Elitzur theorem \cite{Elitzur75}:  any local gauge symmetry cannot be spontaneously broken, because the Elitzur theorem does not apply to the residual local gauge symmetries left after the usual gauge fixing. 
The residual symmetries can be both dependent and independent on spacetime coordinates. 

\noindent

We show that singular topological gauge field configurations play the role of restoring the residual local gauge symmetry violated in the MA gauge \cite{KF21}.
This result implies that color confinement phase is a disordered phase which is realized by non-perturbative effect due to topological configurations. 

As a byproduct, we show that the Abelian U(1) gauge theory in the compact formulation can confine electric charges even in $D=4$ specetime dimensions as discussed long ago by Polyakov \cite{Polyakov75}  in the phase where topological objects recover the residual local gauge symmetry.

\noindent

\section{The residual gauge symmetry in Abelian gauge theory}

Consider QED, or any local U(1) gauge-invariant system with the total Lagrangian density
\begin{align}
\mathscr{L} &= \mathscr{L}_{\text{inv}} + \mathscr{L}_{\text{GF+FP}} .
\end{align}
Here the gauge-invariant part $\mathscr{L}_{\text{inv}}$ is invariant under the local gauge transformation:
\begin{align}
  A_\mu(x) \to A_\mu^\omega(x) := A_\mu(x)+\partial_\mu \omega(x) .
\end{align}
To fix this gauge degrees of freedom, we introduce the Lorenz gauge fixing condition:
\begin{align}
  \partial^\mu A_\mu(x)=0 .
\end{align}
Then the gauge-fixing (GF) and the Faddeev-Popov (FP) ghost term is given by
\begin{align}
  \mathscr{L}_{\text{GF+FP}} 
 = 
  B \partial^\mu A_\mu(x) 
  + \frac{1}{2} \alpha {B}^2 - i \partial^\mu \bar{{c}} \partial_\mu {c} .
\end{align}
However, this gauge-fixing still leaves the invariance under the transformation  function 
 $\omega(x)$ linear in $x^\mu$:
\begin{align}
 \omega(x)  =  a+  \epsilon_\rho x^\rho  ,
\end{align}
since this is a solution of the equation:
\begin{align}
\partial^\mu \partial_\mu \omega(x)=0 \Longrightarrow 
   \partial^\mu A_\mu^\omega(x) = \partial^\mu A_\mu(x)+\partial^\mu \partial_\mu \omega(x)  =0  .
\end{align}
This symmetry is an example of the residual local gauge symmetry. 

There are two conserved charges, the usual charge $Q$ and the vector charge $Q^\mu$, as generators of the transformation:
\begin{align}
 \delta^\omega A_\mu(x):=A_\mu^\omega(x) - A_\mu(x)
 =  [i(aQ+ \epsilon_\rho Q^\rho  ), A_\mu(x)] 
 =  \partial_\mu \omega(x) = \epsilon_\mu . 
\end{align}
This relation must hold for arbitrary $x$-independent constants $a$ and $\epsilon_\mu$, leading to the commutator relations: 
\begin{align}
  [i Q , A_\mu(x)] = 0, \quad 
[i Q^\rho , A_\mu(x)] = \delta^\rho_\mu . 
\end{align}
The first equation implies that the usual $Q$ symmetry, i.e., the global gauge symmetry is  not spontaneously broken:
\begin{align}
 \langle 0|[i Q  , A_\mu(x)]|0 \rangle = 0 ,
\end{align}
while the second equation implies that $Q^\mu$ symmetry, i.e., the residual local gauge symmetry is  always spontaneously broken:
\begin{align}
 \langle 0|[i Q^\rho , A_\mu(x)]|0 \rangle = \delta_\mu^\rho .
\end{align}
Ferrari and Picasso \cite{FP71} argued from this observation that photon is understood as the massless Nambu-Goldstone (NG) vector boson associated with the spontaneous breaking of $Q^\mu$ symmetry according to the Nambu-Goldstone theorem. 
See e.g., \cite{KTU85} for more details. 
Anyway, the restoration of the residual local gauge symmetry does not occur in the ordinary  Abelian case.

\section{Color confinement and residual local gauge symmetry}

First of all, we recall the result of Kugo and Ojima on color confinement.

\textbf{Proposition 1}:
[Kugo-Ojima color confinement criterion (1979)]  \cite{KO79}
Choose the Lorenz gauge fixing $\partial^\mu \mathscr{A}_\mu=0$. 
Suppose that the BRST symmetry exists.
Let  $\mathcal{V}_{\rm phys}$ be the physical state space with 
$\langle {\rm phys} |  {\rm phys}\rangle \ge 0$ as a subspace of an indefinite metric state space 
$\mathcal{V}$ defined by the BRST charge operator $Q_{\rm B}$  as
\begin{align}
 \mathcal{V}_{\rm phys} = \{ | {\rm phys}\rangle \in  \mathcal{V};  Q_{\rm B} |{\rm phys}\rangle=0 \} \subset   \mathcal{V} .
\label{KO-cond}
\end{align}
Introduce the function $u^{AB}(p^2)$ called the \textbf{Kugo-Ojima (KO) function} defined by
\begin{align}
 u^{AB}(p^2) \left( g_{\mu\nu} - \frac{p_\mu p_\nu}{p^2} \right) =  \int d^Dx \ e^{ip(x-y)} \langle 0 | {\rm T}^{}[(\mathscr{D}_\mu\mathscr{C})^A (x) g(\mathscr{A}_\mu \times \bar{\mathscr{C}})^B(y) | 0 \rangle .
\label{color-KO-f}
\end{align}
If the condition called \textbf{Kugo-Ojima (KO) color confinement criterion}  is satisfied in the Lorenz gauge 
\begin{align}
  \lim_{p^2 \to 0} u^{AB}(p^2) = -\delta^{AB} ,
\label{CCF-KO-c}
\end{align}
then the color charge operator $Q^A$ is well defined, namely, the color symmetry is not spontaneously broken, and 
  $Q^A$ vanishes for any physical state $ \Phi, \Psi \in \mathcal{V}_{\rm phys}$,
  \begin{align}
 \langle \Phi | Q^{A} | \Psi \rangle=0 , \quad \Phi, \Psi \in \mathcal{V}_{\rm phys} .
\label{color-KO2}
\end{align}
The BRST singlets as physical particles are all color singlets, while colored particles belong to the BRST quartet representation. 
Therefore, all colored particles cannot be observed and only color singlet particles can be observed.  

Hata \cite{Hata82} investigated the possibility of the restoration of the residual  \textbf{``local gauge symmetry''} in non-Abelian gauge theories with covariant gauge fixing, which is broken in perturbation theory due to the presence of massless gauge bosons even when the \textbf{global gauge symmetry} is unbroken. 
Note that ``local gauge symmetry'' with the quotation marks means that it is not exactly conserved, but is conserved only in the physical subspace $\mathcal{V}_{\rm phys}$ of the state vector space $\mathcal{V}$.

\textbf{Proposition 2}:
[Hata (1982)]  \cite{Hata82}
Consider the residual ``local gauge symmetry'' specified by 
$\omega(x) \in su(N) $ linear in $x^\mu$:
\begin{align}
 \omega(x)  = T_A \omega^A(x) , \  \omega^A(x) = \epsilon_\rho^A x^\rho ,
\label{color-Hata1}
\end{align}
where $\epsilon_\rho^A$ is $x$-independent constant parameters. 
Then there exists the Noether current 
\begin{align}
 \mathscr{J}^\mu_\omega(x)  = gJ^\mu{}^A(x)  x^\rho \epsilon_\rho^A + \mathscr{F}^{\mu\rho}{}^A(x) \epsilon_\rho^A := \mathscr{J}^\mu{}_\rho^A (x) \epsilon^\rho{}^A ,
\label{color-Hata1b}
\end{align}
which is conserved only in the physical subspace $\mathcal{V}_{\rm phys}$ of the state vector space $\mathcal{V}$: 
\begin{align}
 \langle \Phi | \partial_\mu \mathscr{J}^\mu_\omega(x) | \Psi \rangle=0 , \quad \Phi, \Psi \in \mathcal{V}_{\rm phys} ,
\label{color-KO2}
\end{align}
where $J^\mu{}^A(x)$ is the Noether current  associated with the global gauge symmetry which is conserved in 
 $\mathcal{V}$. 
Then the Ward-Takahashi (WT) relation holds for the local gauge current
 $\mathscr{J}^\mu{}_\rho^A (x) $ communicating to 
 $\mathscr{A}_\sigma^B(y)$:
\begin{align}
 \int d^Dx \ e^{ip(x-y)} \partial_\mu^x \langle 0 | {\rm T}^{}[\mathscr{J}^\mu{}_\rho^A (x)  \mathscr{A}_\sigma^B(y)] | 0 \rangle 
 = i \left( g_{\rho\sigma} - \frac{p_\rho p_\sigma}{p^2} \right) [\delta^{AB}+u^{AB}(p^2)] .
\label{color-Hata1}
\end{align}
Thus, if the KO condition  in the Lorenz gauge is satisfied
\begin{align}
  \lim_{p^2 \to 0} u^{AB}(p^2) = -\delta^{AB} ,
\label{color-KO-c2}
\end{align}
then the massless ``Nambu-Goldstone pole'' between 
 $\mathscr{J}^\mu{}_\rho^A$ and 
  $\mathscr{A}_\sigma^B$ contained in perturbation theory disappears.  
\\
The restoration condition coincides exactly with the Kugo and Ojima color confinement criterion! 
This means that the residual local gauge symmetry is restored if the KO condition is satisfied. 

We define the \textbf{restoration of the residual ``local gauge symmetry''} as the \textbf{disappearance of the massless ``Nambu-Goldstone pole'' } from the local gauge current $\mathscr{J}^\mu{}_\rho^A (x)$ communicating to the gauge field $\mathscr{A}_\sigma^B(y)$ through the WT relation.
In this sense, quarks and other colored particles are shown to be confined in the local gauge symmetry restored phase.

\section{Residual gauge symmetry in the Lorenz gauge}

The total Lagrangian density is given by
\begin{align}
\mathscr{L} &= \mathscr{L}_{\text{inv}} + \mathscr{L}_{\text{GF+FP}} .
\label{eq:QCD-Lagrangian}
\end{align}
The first term $\mathscr{L}_{\text{inv}}$ is the gauge-invariant part for the gauge field $\mathscr{A}_\mu$ and the matter field $\varphi$ given by
\begin{align}
  \mathscr{L}_{\text{inv}} &= - \frac{1}{4} \mathscr{F}_{\mu \nu} \cdot \mathscr{F}^{\mu \nu} + \mathscr{L}_{\text{matter}} (\psi , D_\mu \psi) ,
\end{align}
with 
$
  \mathscr{F}_{\mu \nu} := \partial_\mu \mathscr{A}_\nu - \partial_\nu \mathscr{A}_\mu + g \mathscr{A}_\mu \times \mathscr{A}_\nu = - \mathscr{F}_{\nu \mu} 
$
and  
$D_\mu \psi:=\partial_\mu \psi -ig\mathscr{A}_\mu \psi$.
\\
The second term $\mathscr{L}_{\text{GF+FP}}$  is the sum of the the gauge-fixing (GF) term and the Faddeev-Popov (FP) ghost term 
where the GF term includes the Nakanishi-Lautrup field $\mathscr{B} (x)$ which is the Lagrange multiplier field to incorporate the gauge fixing condition and the FP ghost term includes the ghost field $\mathscr{C}$ and the antighost field $\bar{\mathscr{C}}$. 

For the gauge field and the matter field, we consider  the local gauge transformation with the Lie algebra-valued transformation function $\omega (x)=\omega^A(x)T_A$  given by
\begin{align}
  \delta^{\omega} \mathscr{A}_\mu (x) &= \mathscr{D}_\mu \omega (x) := \partial_\mu \omega (x)  + g\mathscr{A}_\mu \times \omega (x) , \nonumber\\
  \delta^\omega \varphi (x) &= i g \omega (x) \varphi (x) ,
\nonumber\\
    \delta^{\omega} \mathscr{B} (x) &= g \mathscr{B} (x) \times \omega (x) , \nonumber\\
  \delta^{\omega} \mathscr{C} (x) &= g \mathscr{C} (x) \times \omega (x) , \nonumber\\
  \delta^{\omega} \bar{\mathscr{C}} (x) &= g \bar{\mathscr{C}} (x) \times \omega (x) .
  \label{eq:nonabelian-gauge-transformation}
\end{align}
Now we proceed to write down the Ward-Takahashi relation to examine the appearance or disappearance of the massless  ``Nambu-Goldstone pole''. 
We consider the condition for the restoration of the residual local gauge symmetry for a general $\omega$. 
We focus on the WT relation 
\begin{align}
 & \int d^D x e^{i p (x - y)} \partial_\mu^x \braket{ {\rm T}^{} \mathscr{J}_\omega^\mu (x) \mathscr{A}_\lambda^B (y)} \nonumber\\
  = & i 
   \braket{\delta^\omega \mathscr{A}_\lambda^B(y)} 
  + \int d^D x \  e^{i p (x - y)} \braket{ {\rm T}^{} \partial_\mu \mathscr{J}_\omega^\mu (x) \mathscr{A}_\lambda^B (y)}   \nonumber\\
  = & i \braket{ \partial_\lambda \omega^B(y) + g(\mathscr{A}_\lambda \times \omega)^B(y) } + \int d^D x e^{i p (x - y)} \braket{ {\rm T}^{}\delta^\omega \mathscr{L}_{\rm GF+FP}(x)  \mathscr{A}_\lambda^B(y)} \nonumber\\
  = &  i \partial_\lambda \omega^B (y) + \int d^D x e^{i p (x - y)} \braket{ {\rm T}^{}\delta^\omega \mathscr{L}_{\rm GF+FP}(x)  \mathscr{A}_\lambda^B(y)}    ,
  \label{CCF-WTI_gen1}
\end{align}
where we have assumed the unbroken Lorentz invariance to use $\braket{0 | \mathscr{A}_\lambda (x) | 0} = 0$  in the final step. 
Note that this relation is valid for any choice of the gauge fixing condition. 

For the Lorenz gauge $\partial_\mu \mathscr{A}^\mu=0$, the GF+FP term is given by
\begin{align}
  \mathscr{L}_{\text{GF+FP}} 
 =  
  \mathscr{B} \cdot \partial_\mu \mathscr{A}^\mu 
  + \frac{1}{2} \alpha \mathscr{B} \cdot \mathscr{B} - i \partial^\mu \bar{\mathscr{C}} \cdot \mathscr{D}_\mu \mathscr{C} 
 = - i {\boldsymbol \delta}_{\rm{B}}  \left[ \bar{\mathscr{C}} \cdot  \left( \partial^\mu \mathscr{A}_\mu + \frac{\alpha}{2} \mathscr{B} \right) \right] ,
  \label{eq:Lorenz-GF+FP}
\end{align}
where $\alpha$ is the gauge-fixing parameter.
The change under the generalized local gauge transformation is given by $\alpha$-independent expression:
\begin{align}
 		\delta^\omega \mathscr{L}_{\rm GF+FP}(x)
		 = i  \bm{\delta}_{\rm B} (\mathscr{D}_\mu \bar{\mathscr{C}}(x))^A  \partial^\mu \omega^A(x) .
\end{align}
In the Lorenz gauge, the above WT relation (\ref{CCF-WTI_gen1}) reduces to 
\begin{align}
  &\int d^D x e^{i p (x - y)} \partial^x_\mu \braket{ {\rm T}^{} {\mathscr{J}_\omega^{\mu }}_\nu^A (x) \partial^\nu \omega^A(x) \mathscr{A}_\lambda^B (y)}  
  \nonumber\\
  = & i \partial_\lambda \omega^B (y) + \int d^D x e^{i p (x - y)} \partial^\mu \omega^A(x) \braket{ {\rm T}^{} i  \bm{\delta}_B (\mathscr{D}_\mu \bar{\mathscr{C}}(x))^A \mathscr{A}_\lambda^B(y)} .
  \label{CCF-WTI_gen_Lorenz}
\end{align}
The second term of \eqref{CCF-WTI_gen_Lorenz}  is rewritten using 
$\bm{\delta}_B( \mathscr{D}_\mu \bar{\mathscr{C}})=\bm{\delta}_B( \partial_\mu \bar{\mathscr{C}} + g(\mathscr{A}_\mu \times \bar{\mathscr{C}}) )
=- \partial_\mu \mathscr{B}+  g\bm{\delta}_B(\mathscr{A}_\mu \times \bar{\mathscr{C}})$ 
\begin{align}
  &\int d^D x e^{i p (x - y)} \partial^\mu \omega^A(x) \braket{ {\rm T}^{} i    \bm{\delta}_B (\mathscr{D}_\mu \bar{\mathscr{C}}(x))^A \mathscr{A}_\lambda^B (y)} \nonumber\\
  = &-  \int d^D x e^{i p (x - y)} \partial^\mu \omega^A(x) \partial_\mu^x  i \frac{\partial_{\lambda}^x }{\partial_x^2} \delta^D(x-y) \delta^{AB}  \nonumber\\
  &+ i \int d^D x e^{i p (x - y)} \partial^\mu \omega^A (x)
  \left(g_{\mu\lambda}- \frac{\partial_{\mu}^x \partial_{\lambda}^x}{\partial_x^2} \right) u^{AB}(x-y) ,
  \label{eq:WI_Lorenz_general_2nd}
\end{align}
where we have used $\braket{\bm{\delta}_{\rm B} F} = 0$ for any functional $F$ due to the physical state condition,  
the exact form of the propagator in the Lorenz gauge
\begin{align}
 \langle 0 | {\rm T}^{} \mathscr{A}_{\mu}^A(x) \mathscr{B}^B(y) | 0 \rangle=
\langle 0 |{\rm T}^{*} (\mathscr{D}_{\mu} \mathscr{C})^A(x) i \bar{\mathscr{C}}^B(y) | 0 \rangle
=&  i \frac{\partial_{\mu}^x }{\partial_x^2} \delta^D(x-y) \delta^{AB} ,
\label{CCF-DCC00}
\end{align}
and 
the definition of the \textbf{Kugo-Ojima (KO) function}\index{Kugo-Ojima (KO) function} $u^{AB}$ in the configuration space
\begin{align}
\langle 0 | {\rm T}^{} (\mathscr{D}_{\mu} \mathscr{C})^A(x) (g \mathscr{A}_{\nu} \times \bar{\mathscr{C}})^B(y) | 0 \rangle
= \left(g_{\mu\nu}- \frac{\partial_{\mu}^x \partial_{\nu}^x}{\partial_x^2} \right) u^{AB}(x-y) .
\label{CCF-KO-func-real}
\end{align}
Thus, we obtain the general condition in the Lorenz gauge written in the Euclidean form:
\begin{align}
\boxed{
  \lim_{p \rightarrow \ 0}  \int d^D x e^{i p (x - y)} \partial_\mu \omega^A (x) \left(\delta_{\mu \lambda} -  \frac{\partial_{\mu}^x \partial_{\lambda}^x}{\partial_x^2} \right)  
   \left[ \delta^D(x-y)\delta^{AB} + u^{AB}(x-y) \right] = 0 },
  \label{eq:condition_Lorenz_general}
\end{align}
This confinement criterion can be applied to the Abelian and non-Abelian gauge theory as well irrespective of the compact or non-compact formulation, 
and is able to understand confinement in all the cases. 

In the non-compact gauge theory formulated in terms of the Lie-algebra-valued gauge field, the choice of $\omega^A (x)$ as the non-compact variable linear in $x$,
\begin{align}
\omega^A (x) = \text{const.} + \epsilon_\mu^A x_\mu =\text{const. + non-compact~variable} ,
\end{align}
is allowed. 
Indeed, for this choice, the criterion (\ref{eq:condition_Lorenz_general}) is reduced to
\begin{align}
  \epsilon_\mu^A \lim_{p \rightarrow \ 0}  \left( \delta_{\mu \lambda} - \frac{p_\mu p_\lambda}{p^2} \right) 
 \left[ \delta^{AB}  + \tilde u^{AB} (p) \right] = 0.
  \label{eq:condition_Lorenz_special}
\end{align}
This reproduces the KO condition $\tilde u^{AB}(0) = -\delta^{AB}$ as first shown by Hata.

For the Abelian gauge theory, the KO function is identically zero $u^{AB}(x) \equiv 0$, i.e., $\tilde u^{AB}(0)=0$. Therefore, the KO condition is not satisfied, which means no confinement in the Abelian gauge theory.

In the compact gauge theory, however, confinement does occur even in the Abelian gauge theory, as is well known in the lattice gauge theory. 
This case is also understood by the above criterion.

\section{Restoration of residual local symmetry  in MA gauge}

We decompose the Lie-algebra valued quantity to the diagonal Cartan part and the remainig off-diagonal part, e.g., the gauge field $\mathscr{A}_\mu=\mathscr{A}_\mu^A T_A$ with the generators $T_A$ ($A=1,\ldots,N^2-1$) of the Lie algebra $su(N)$ has the decomposition:
\begin{align}
  \mathscr{A}_\mu(x)=\mathscr{A}_\mu^A(x) T_A
  = a_\mu^j(x) H_j + A_\mu^a(x) T_a ,
  \label{eq:decomp}
\end{align}
where $H_j$ are the Cartan generators and $T_a$ are the remaining generators of the Lie algebra $su(N)$. 
In what follows, the indices $j,k,\ell,\ldots$ label the diagonal components and the indices $a,b,c,\ldots$ label  the off-diagonal components. 
The maximal Abelian (MA) gauge is given by
\begin{align}
  (\mathscr{D}_{}^\mu [a] A_\mu(x))^a := \partial^\mu A_\mu^a(x) + {} gf^{ajb}a^\mu{}^j(x) A_\mu^b(x)
  = 0 .
  \label{eq:MA-condition}
\end{align}
The MA gauge is a partial gauge which fix the off-diagonal components, but does not fix the diagonal components.  Therefore, we further  impose the Lorenz gauge for the diagonal components
\begin{align}
  \partial^\mu a_\mu^j(x) = 0 .
  \label{eq:MA-Lorenz}
\end{align}
The GF+FP term for the gauge-fixing condition \eqref{eq:MA-condition} and \eqref{eq:MA-Lorenz} 
 is given 
using the BRST transformation as 
\begin{align}
  \mathscr{L}_{\text{GF+FP}} 
  = &- i \bm{\delta}_{\rm B}\left\{ \bar{C}^a \left(\mathscr{D}_{}^\mu [a] A_\mu + \frac{\alpha}{2} B\right)^a \right\} 
  - i \bm{\delta}_{\rm B} \left\{ \bar{c}^j \left(\partial^\mu a_\mu + \frac{\beta}{2} b\right)^j \right\} ,
  \label{eq:MA-GF+FP1}
\end{align}
which reads
\begin{align}
  \mathscr{L}_{\text{GF+FP}} = &- (\mathscr{D}_{}^\mu [a]^{b a} B^a) A_\mu^b + \frac{\alpha}{2} B^a B^a - i (\mathscr{D}_{}^\mu [a]^{b a} \bar{C}^a) \mathscr{D}_\mu [a]^{b c} C^c \nonumber\\
  &- i g (\mathscr{D}_{}^\mu [a]^{b a} \bar{C}^a) f^{b c d} A_\mu^c C^d - i g (\mathscr{D}_{}^\mu [a]^{b a} \bar{C}^a) f^{b c j} A_\mu^c c^j \nonumber\\
  &+ i {} g \bar{C}^a f^{a j b} \partial_\mu c^j A^{\mu b} + i {} g^2 \bar{C}^a f^{a j b} f^{j c d} A_\mu^c C^d A^{\mu b}  \nonumber\\
  &- \partial^\mu b^j a_\mu^j+ \frac{\beta}{2} b^j b^j - i \partial^\mu \bar{c}^j \partial_\mu c^j - i g \partial^\mu \bar{c}^j f^{j a b} A_\mu^a C^b .
  \label{eq:MA-GF+FP2}
\end{align}
The local gauge transformation of the Lagrangian has the following form 
\begin{align}
		\delta^\omega \mathscr{L} &= \delta^\omega \mathscr{L}_{\text{GF+FP}}  
		= \partial_\mu \mathscr{J}^\mu_\omega 
 = 
 g \partial_{\mu} \mathscr{J}^{\mu}  \cdot \omega 
+ \left[    \partial_{\nu} \mathscr{F}^{\mu \nu}   + g \mathscr{J}^{\mu}  \right] \cdot \partial_{\mu} \omega 
\nonumber\\  
&= 
 g \partial^\mu {J}_{\mu}^j \omega^j 
+ \left[ \partial^{\nu} {f}_{\mu \nu}^j + g {J}_{\mu}^j \right]  \partial_{\mu} \omega^j
+ g \partial^\mu {J}_{\mu}^a \omega^a  
+ \left[ \partial^{\nu} {F}_{\mu \nu}^a + g {J}_{\mu}^a \right]  \partial_{\mu} \omega^a
\nonumber\\  
&= 
i \bm{\delta}_B \partial_\mu \bar{c}^j \partial^\mu \omega^j
+ i {\boldsymbol \delta}_{\rm{B}} \partial^\mu (\mathscr{D}_{\mu}[\mathscr{A}]\bar{\mathscr{C}})^a \omega^a  
+ i {\boldsymbol \delta}_{\rm{B}} (\mathscr{D}_{\mu}[\mathscr{A}]\bar{\mathscr{C}})^a \partial^\mu \omega^a  
.
		\label{eq:MA-current-divergence2}
\end{align}
This is BRST exact,  showing that the local gauge current $\mathscr{J}^\mu_\omega$  is conserved in the physical state space. 
\\
The WT relation in the MA gauge can be calculated in the similar way to the Lorenz gauge by using \eqref{eq:MA-current-divergence2} as follows.
We focus on the diagonal gauge field $a_\lambda^k$. 
Consequently, we obtain the condition  for the restoration of the residual local gauge symmetry for the diagonal gauge field \cite{KF21}
\begin{align}
& \lim_{p \rightarrow \ 0} \int d^D x \ e^{i p (x - y)} \partial_\mu^x \braket{ {\rm T}^{} \mathscr{J}_\omega^\mu (x) {a}_\lambda^k (y)}  \nonumber\\
&=
\boxed{  \lim_{p \rightarrow \ 0}  i \int d^D x \ e^{i p (x - y)}  \partial^\mu \omega^k(x) 
  ( \delta_{\mu\lambda} \Box_D-\partial_\mu \partial_\lambda) \Box_D^{- 1} (x , y)   = 0 } ,
  \label{eq:condition_MAG_diag_general}
\end{align}
where $\Box_D^{- 1} (x , y)$ denotes the Green function of the  Laplacian $\Box_D=\partial_\mu \partial_\mu$ in the $D$-dimensional Euclidean space.
\\
If we choose $\omega^j(x) = \epsilon_\nu^j x^\nu$, this indeed reproduces non-vanishing divergent result.  
\\
However, this choice must be excluded in the MA gauge, since the maximal torus subgroup $U(1)^{N-1}$  for the diagonal components is a compact subgroup of the compact $SU(N)$ group. 
 In some sense, $\omega^j(x)$ must be angle variables reflecting the compactness of the gauge group. 
 
In the compact gauge theory formulated in terms of the group-valued gauge field, on the other hand, we must choose the compact, namely, angle variables for $\omega^A$, 
\begin{align}
\omega^A (x) =\text{const. + angle~variable}=\text{const. + compact~variable} .
\end{align}

For concreteness, we consider the $SU(2)$ case with singular configurations  coming from the angle variables. 
In what follows, we work in the Euclidean space and use subscripts instead of the Lorentz indices.  
As the residual gauge transformation, we take the following examples which satisfy both the Lorenz gauge condition $\partial_\mu \mathscr{A}_\mu^A=0$ and the MA gauge condition $(\mathscr{D}_\mu [a] A_\mu)^a =0$ (and $\partial^\mu a_\mu^j=0$). 
\\
\noindent
$\bullet$
For $D=2$, a collection of vortices of Abrikosov-Nielsen-Olesen type (1979) \cite{NO79}
\begin{align}
   \partial_\mu \omega^j (x)
   = \sum_{s=1}^{n} C_s \varepsilon_{j\mu\nu} \frac{(x-a_s)_\nu}{|x-a_s|^2} \ (j=3, \ \mu,\nu=1,2)  \ (x,a_s \in \mathbb{R}^2) ,
   \label{sing-2}
\end{align}
where $C_s$  ($s=1,\ldots,n$) are  arbitrary constants.
This type of $\omega(x)$ is indeed an angle variable $\theta$ going around a point $a=(a_1, a_2) \in \mathbb{R}^2$, because 
\begin{align}
   \omega (x) = \theta(x) 
   =: \arctan \frac{x_2-a_2}{x_1-a_1} 
   \Longrightarrow 
   \partial_\mu \omega (x) 
   = - \varepsilon_{\mu\nu} \frac{x_\nu-a_\nu}{(x_1-a_1)^2+(x_2-a_2)^2} \ (\mu=1,2).
   \label{sing-2b}
\end{align}
This is a topological configuration which is classified by the winding number of the map from the circle in the space to the circle in the target space: $S^1 \to U(1) \cong S^1$, i.e., by the first Homotopy group $\pi_1(S^1)=\mathbb{Z}$. 
\\
\noindent
$\bullet$
For $D=3$, a collection of magnetic monopoles of the Wu-Yang type (1975) \cite{WY75}, 
%
which corresponds to the zero size limit of the `t Hooft-Polyakov magnetic monopole (1974) \cite{tHP74}
%
\begin{align}
   \partial_\mu \omega^j (x)
   = \sum_{s=1}^{n}  C_s  \varepsilon_{j\mu\nu} \frac{(x-a_s)_\nu}{|x-a_s|^2} \ (j=3, \ \mu,\nu=1,2,3)  \ (x,a_s \in \mathbb{R}^3) .
   \label{sing-3}
\end{align}
A magnetic monopole is a topological configuration which is classified by the winding number of the map from the sphere in the space to the sphere in the target space: $S^2 \to SU(2)/U(1) \cong S^2$, i.e., by the second Homotopy group $\pi_2(S^2)=\mathbb{Z}$. 
\\
\noindent
$\bullet$ For $D=4$,  a collection of merons of Alfaro-Fubini-Furlan (1976) \cite{AFF76}, 
instantons of the Belavin-Polyakov-Shwarts-Tyupkin (BPST) type  (1975)  \cite{BPST75} 
in the non-singular gauge with zero size, 
\begin{align}
   \partial_\mu \omega^j (x)
   = \sum_{s=1}^{n}  C_s \eta^{j}_{\mu\nu} \frac{(x-a_s)_\nu}{|x-a_s|^2} \ (j=3, \ \mu,\nu=1,2,3,4)  \ (x,a_s \in \mathbb{R}^4) .
   \label{sing-4}
\end{align}
Meron and instanton are  topological configuration which are classified by the winding number of the map from the 3-dimensional sphere in the space to the sphere in the target space: $S^3 \to SU(2) \cong S^3$, i.e., by the third Homotopy group $\pi_3(S^3)=\mathbb{Z}$. 
%

By taking into account 
 $\varepsilon^{j}_{\mu\nu}=-\varepsilon^{j}_{\nu\mu}$, $\eta^{j}_{\mu\nu}=-\eta^{j}_{\nu\mu}$, it is easy to show that all these configurations satisfy the Laplace equation $\Box \omega^j(x)=0$ almost everywhere except for the locations  $a_s \in \mathbb{R}^D$
 of the singularities: $\Box \omega^j(x)= \sum_{s=1}^{n} C_s  \delta^D(x-a_s)$. 
These configurations are examples of the classical solutions of the Yang-Mills field equation with non-trivial topology.

We can show that the restoration condition is satisfied for these singular configurations \cite{KF21}:
\begin{align}
\boxed{  \lim_{p \to 0}   \int d^D x \ e^{i p (x - y)} \frac{(x-a_s)_\nu}{|x-a_s|^2}  \left( \delta_{\mu\lambda} \Box_D -\partial_\mu \partial_\lambda  \right)   \frac{\frac{\Gamma \left(\frac{D}{2} - 1\right)}{4 \pi^{D / 2}}}{(|x - y|^2)^{\frac{D - 2}{2}}}
  =0 } .
  \label{eq:MA-WI-singular2}
\end{align}
where we have used the expression of  the Green function $\Box_D^{- 1} (x , y)$  of the  Laplacian $\Box_D=\partial_\mu \partial_\mu$ in the $D$-dimensional Euclidean space  given by
\begin{align}
  \Box^{- 1}_D (x , y) = \int \frac{d^D p}{(2 \pi)^D} e^{i p (x - y)} \frac{1}{- p^2} = - \frac{\Gamma \left(\frac{D}{2} - 1\right)}{4 \pi^{D / 2}} \frac{1}{|x - y|^{D - 2}} ,
  \label{Green-Laplacian}
\end{align}
where $\Gamma$ is the gamma function with the integral representation given by 
\begin{align}
  \Gamma (z) = \int_0^\infty dt \ t^{z - 1} e^{- t} \ (z > 0) .
\end{align}
For any $D \ge 2$,  this integral (\ref{eq:MA-WI-singular2})  goes to zero linearly in $p$ in the limit $p \to 0$ \cite{KF21}. 
Therefore, the restoration of the residual local gauge symmetry occurs. 

\section{Conclusion and discussion}

\noindent
$\rhd$ Conclusions: we summarize our results:

\noindent
$\bullet$
We have reexamined the restoration of the residual local gauge symmetry left even after imposing the gauge fixing condition in quantum gauge field theories. 
This leads to a generalization of the color confinement criterion.


\noindent
$\bullet$
We have found an important lesson to understand color confinement in quantum gauge theories that the compactness and non-compactness must be discriminated for the gauge transformation of the gauge field. 

\noindent
$\bullet$
The Kugo-Ojima color confinement  criterion can be applied only to the non-compact gauge theory. 
This is a reason why the Kugo-Ojima criterion obtained in the Lorenz gauge cannot be applied to the Maximal Abelian gauge (maximal torus group is a compact group).

\noindent
$\bullet$
In the Maximal Abelian gauge we have shown that the restoration of the residual local gauge symmety  indeed occurs  for the SU(N) Yang-Mills theory in two-, three- and four-dimensional Euclidan spacetime once the singular topological configurations of gauge fields are taken into account. 

\noindent
$\bullet$
This result indicates that the color confinement phase is a disordered phase caused by non-trivial topological configurations irrespective of the gauge choice.

\noindent
$\bullet$
As a byproduct, we find that the compact U(1) gauge theory can have the disordered confinement phase, while the non-compact U(1) gauge theory has the deconfined Coulomb phase. 

\noindent
$\rhd$ Future perspectives: we have the issues to be investigated in future:

\noindent
$\bullet$
Gribov copies, existence of BRST symmetry, 

\noindent
$\bullet$
Higgs phase, Brount-Englert-Higgs (BEH) mechanism, 

\noindent
$\bullet$
Finite temperatures,






\section*{Acknowledgements}


This work was  supported by Grant-in-Aid for Scientific Research, JSPS KAKENHI Grant Number (C) No.19K03840.

\end{document}